# Disability for HIV and Disincentives for Health: The Impact of South Africa's Disability Grant on HIV/AIDS Recovery

Noah Haber,[1,2,3] Till Bärnighausen,[1,2,4] Jacob Bor,[2,5,6] Jessica Cohen,[1] Frank Tanser,[2,7] Deenan Pillay,[2,8] Günther Fink[1,9]

1 Department of Global Health and Population, Harvard T.H. Chan School of Public Health, Boston, USA
2 Africa Health Research Institute, Somkhele, South Africa
3 Carolina Population Center, University of North Carolina at Chapel Hill, Chapel Hill, USA
4 Institute of Public Health, University of Heidelberg, Heidelberg, Germany
5 Department of Global Health, Boston University School of Public Health, Boston, USA
6 Health Economics and Epidemiology Research Office, University of the Witwatersrand, Johannesburg, South Africa
7 School of Nursing and Public Health, University of KwaZulu-Natal, Durban, South Africa
8 Division of Infection and Immunity, University College London, UK
9 Swiss Tropical and Public Health Institute, University of Basel, Basel, Switzerland

Corresponding author:
Noah Haber
nhaber@unc.edu

# Abstract

South Africa's disability grants program is tied to its HIV/AIDS recovery program, such that individuals who are ill enough may qualify. Qualification is historically tied to a CD4 count of 200 cells/µL, which improve when a person adheres to antiretroviral therapy. This creates a potential unintended consequence where poor individuals, faced with potential loss of their income, may choose to limit their recovery through non-adherence. To test for manipulation caused by grant rules, we identify differences in disability grant recipients and non-recipients' rate of CD4 recovery around the qualification threshold, implemented as a fixed-effects difference-in-difference around the threshold. We use data from the Africa Health Research Institute Demographic and Health Surveillance System (AHRI DSS) in rural KwaZulu-Natal, South Africa, utilizing DG status and laboratory CD4 count records for 8,497 individuals to test whether there are any systematic differences in CD4 recover rates among eligible patients. We find that disability grant threshold rules caused recipients to have a relatively slower CD4 recovery rate of about 20-30 cells/µL/year, or a 20% reduction in the speed of recovery around the threshold.

# Introduction

While a large literature has highlighted the effects of income or means-tested welfare programs on labor supply, relatively little is known regarding the impact of these programs on health. South Africa's disability grant (DG) program may be an example where a program designed to improve economic wellbeing can have a negative impact on health. The DG program in South Africa includes qualification for people with HIV/AIDS by having a low CD4 count, a biological marker for illness due to HIV/AIDS. Because CD4 counts improve with treatment, individuals relying on grants face a potential incentive to stop treatment to remain qualified for the grant. This paper utilizes an unusual identification strategy to identify the existence of this potential unintended consequence on HIV-related health and behavior.

*South Africa's disability grant program and HIV/AIDS*

South Africa's disability grants system provides income for those who are unable to work for health reasons. The grants make up a substantial portion of income, and may increase self-reported quality of life [1-3]. In 2012, the grant size was R1,200 (approximately US$118) per month [4], comprising 41-49% of total household income for grant recipient households in 2002 in Cape Town [1]. These grants cover six month blocks, after which participants need to re-apply for a new grant block [2, 5]. In extremely poor regions, such as rural KwaZulu-Natal, South Africa, qualitative evidence from Phaswana-Mafuya, Peltzer [2] shows that this source of income may be critical for households which may have to make the choice between food and medicine. Other studies, such as Black, Daniel [6], have shown that disability benefit application is impacted by outside economic conditions and hardship. Given the level of poverty and lack of outside options for income, the incentives to maintain DG status are very strong.

South Africa's disability benefits program is unique in that it explicitly includes disability due to AIDS sickness as a qualifying condition, defined as having a CD4 count of under 200 cells/μL [3, 7]. CD4 lymphocytes are immune cells which are attacked by HIV. CD4 counts are the number of these cells present in one μL of blood as assessed from a laboratory blood test, where lower CD4 counts indicates a more compromised immune system. The CD4 qualification threshold varies in intensity both over time and geography. While laws changed in 2008 to attempt to reduce how strictly CD4 count disability benefits qualification rules bound [4], the threshold of 200 cells/μL is widely understood as the de facto rule for DG qualification due to a combination of precedent, simplicity, and lack of official alternatives to the rule [2, 8].

Effective treatment with antiretroviral therapy (ART) increases CD4 count levels over time [9-11]. If individuals are receiving disability grants due to their low CD4 count, adhering to ART does not only improve their health, but also increases the likelihood of losing the disability grant. Poor or limited adherence suppresses, or even reverses, AIDS-related health recovery, as indicated by CD4 counts [12, 13]. In addition to the direct health threat of non-adherence to the individual, poor adherence also leads to increasing risk of ART resistance [14, 15]. Disability grant recipients may have other ways to maintain eligibility through CD4 count manipulation, including misreporting laboratory results, and or binge-drinking prior to laboratory tests as a strategy to sabotage test results de Paoli, Mills [8].

*Existing literature on disability grant and non-work programs*

Most economics literature on disability and non-work programs focuses almost exclusively on labor supply. These studies are generally based on the US and other high income countries, taking advantage of variation in benefits qualification to estimate the degree to which benefits may reduce labor supply. Autor, Duggan [16] exploits changes in the US Department of Veteran's Affairs eligibility rules and estimate that disability benefits caused an 18 percentage point decrease in labor supply. Exploiting variation in medical examiner assignment, Maestas, Mullen [17] estimate a 28 percentage point reduction in labor supply due to disability benefits in the US. In terms of the impact of disability grants on health behavior, Singleton [18] finds that the US Department of Veteran's Affairs policy change for disability pay increased the likelihood of participants obtaining a diagnosis for diabetes by 2.7 percentage points. At least one study in a developing country, Firpo, Pieri [19], finds evidence that households manipulate their labor supply to qualify for Brazil's *Bolsa Familia*.

Existing literature contains many examples of disability benefit eligibility threshold rules being exploited for causal identification of labor and earnings outcomes. Kostøl and Mogstad [20], Chen and van Der Klaauw [21], and Borghans, Gielen [22] take advantage of age and time based eligibility rules to find substantial impact of disability benefits programs on labor and income. Liebman, Luttmer [23] utilizes five unique thresholds in social security benefits to measure incentive effects on labor supply and income, though they exclude disability benefits from their analysis.

In terms of individual behaviors in response to policy, there is no shortage of evidence for individuals manipulating data and behavior to qualify for government programs. Camacho and Conover [24] show evidence that local governments manipulated reported development scores to qualify for a variety of social welfare programs in Columbia. Courty and Marschke [25] examines gaming the timing of when training organizations report trainee's outcomes in response to training outcome thresholds.

*Identification strategy summary*

This paper explores whether the disability grant qualification rule impacts CD4 recovery rates. If potential disability grant recipients are manipulating their CD4 counts, whether by modifying their ART adherence or by other means, we would expect that as they approach or go over the threshold, they may have some incentive to reduce or cease taking their medications, which leads to a reduction in their CD4 count level on the following visit. We characterize that as a reduction in a time-fixed rate of recovery to neutralize possible differences in time between visits. Given this framework, we would expect that disability grant recipients would have marginally slower CD4 count recovery when their CD4 level is close to the DG qualification threshold due to the incentive to keep CD4 levels low to become or remain qualified.

We identify this effect using a difference-in-difference strategy in recovery rates, where the first difference is whether the individual has received a CD4 test indicating that they are near or just over the CD4 count threshold, and the second difference is whether they are a disability grant recipient. We use data from the Africa Health Research Institute Demographic and Health Surveillance System (AHRI DSS) in rural KwaZulu-Natal, South Africa, utilizing DG status and laboratory CD4 count records for 8,497 individuals. We find that disability grant threshold rules slowed the CD4 recovery rates of recipients by roughly 20-30 cells/μL/year, or a 20% reduction in the speed of recovery around the threshold. This evidence suggests that the CD4 qualification threshold caused manipulation of CD4 recovery.

# Data

The data for this analysis are drawn from AHRI DSS, a large open cohort demographic surveillance site in the Umkanyakude region in KwaZulu-Natal, individually linked to electronic records from the public

HIV clinic and hospital system in the region. The AHRI DSS is a longitudinal surveillance and monitoring mechanism with annual individual and household surveys, linked by individuals, households, and bounded structures, starting in 2003, of individuals aged 15 years and older. This system covers the vast majority of individuals in the 438 km$^2$ region, with a greater than 99% participation rate and data for over 100,000 individuals. The AHRI DSS is linked to a local bioinformatics system known as ARTEMIS, containing complete medical histories, including CD4 counts, for all South African Department of Health clinics and hospitals in the region [26].

The AHRI DSS region is largely rural, with some small semi-urban areas and one urban township, and is one of the poorest regions in South Africa. While the majority of the population lives within the rural areas, most household income is based on wages from employment and social welfare programs, including the disability grants program. HIV prevalence is extremely high in this region, with overall prevalence estimates reaching around 28-29% in 2010 [27]. Our population consists of individuals in the AHRI DSS cohort for whom disability grant history is known at one point in time and who have at least three recorded CD4 count observations (two intervals between CD4 counts).

Disability grants data are gathered from both AHRI DSS and ARTEMIS datasets. AHRI DSS participants were directly asked if they received disability grant income from 2003-2006, and asked if they receive either a disability grant or a care dependency grant from 2007 onward. While we cannot determine whether the recipient received a disability or care dependency grant, we treat them as having ever vs. never received a disability grant. We note that there were only 289 reports of receiving a care dependency grant from 2003-2006, compared to 3,194 reports of disability grants in the greater AHRI DSS region (and approximately 10x the number of DGs as care dependency grants nationally) [28], suggesting that the vast majority of those who answered yes to this question were receiving disability grants. We also utilize disability grants data from ARTEMIS, which contains limited, but largely missing, reports of whether the person was receiving a disability grant at the point of ART initiation.

Because we do not have specific dates of grant recipient status, we utilize a time-invariant dummy for all individuals which equals 1 if the person ever received a disability grant at any time, and 0 otherwise. Persons for whom we have no disability grant status recorded (<1% of persons in the sample) are dropped from the analysis. There several reasons individuals who have a CD4 count below the threshold never receive disability grants. Some don't qualify due to the income requirement or because they are receiving another social grant. Applications and administration is also believed to be difficult, with very long wait periods, lines, and perceived high rates of rejection/misfiling. The threshold itself does not bind strictly, as it is ultimately up to medical providers to decide who qualified, and DGs can be acquired through non-AIDS-related means. Others may simply not report their DG receipt to AHRI DSS. The measurement error in our estimation of receipt of disability grant is likely to bias results towards the null, as there are likely to be many individuals who are receiving DGs but did not receive them, as well as those reporting receiving DGs for reasons other than qualification by AIDS via CD4 counts.

Our main dependent variable is CD4 recovery rate between CD4 counts. Given that observation time intervals vary across individuals, we convert all rates observed into annualized CD4 changes, which is simply the change in value of the CD4 count from one CD4 count to the next in time, divided by the change in time in years between the two CD4 counts.[1] Each observation in our dataset represents an interval between two sequential CD4 counts, with each interval having a CD4 count value and date at the start of the interval, an interval time, and an annualized rate of change during the interval. We omit interval observations with recovery rates in the 1st and 99th percentiles to prevent data errors and outliers from strongly influencing results for all analyses below.

---

[1] $\frac{CD4_{t+1} - CD4_t}{time_{t+1} - time_t}$

# Empirical estimation

## 1.1 Descriptive statistics

Table 1: Descriptive statistics

|  | All patients | DG non-recipients | DG recipients | Difference |
|---|---|---|---|---|
|  | *mean (SD)* | *mean (SD)* | *mean (SD)* | *Difference (p-value)* |
| Age | 32 (13) | 30 (12) | 40 (13) | 10 (<.001) |
| Female | 0.71 (0.45) | 0.72 (0.45) | 0.69 (0.46) | -0.03 (.018) |
| Education years | 7.9 (4.1) | 8.4 (3.9) | 5.5 (4.3) | -2.9 (<.001) |
| Is employed | 0.34 (0.47) | 0.37 (0.48) | 0.23 (0.42) | -0.14 (<.001) |
| Asset index percentile | 52 (27) | 53 (28) | 49 (25) | -4 (<.001) |
| Received DG ever | 0.18 (0.38) | 0 (0) | 1 (0) | 1 (<.001) |
| CD4 cells/µL change between tests | 43 (240) | 43 (246) | 42 (218) | -1 (.762) |
| Days between CD4 counts | 320 (245) | 325 (255) | 301 (204) | -24 (<.001) |
| Annualized CD4 change | 62 (356) | 62 (363) | 61 (331) | -2 (.665) |
|  | n | n | n |  |
| Number of observations | 47,138 | 36,570 | 10,568 |  |
| Individuals | 8,497 | 6,998 | 1,499 |  |

Descriptive statistics specific to individuals (age, gender, education years, employment, and asset index percentile) are taken for each individual only at the first observed CD4 count. If a measure is not made of a given observation on the day of that CD4 count, the value at the nearest date measured is taken. The p-values reported refer to a two-tailed t-test of the difference between disability grant (DG) recipients and non-recipients.

Descriptive statistics for the study population are shown in Table 1. The sample consists of 47,138 CD4 change records across 8,497 individuals. 1,499 of these individuals report to have received a disability grant. Those who have ever received a disability grant are typically older, less educated, less likely to be employed, and had a lower asset index percentile than their non-recipient counterparts. The mean first recorded CD4 count of disability grant recipients was non-significantly lower than for non-recipients. CD4 counts were measured in intervals of 320 days on average, with mean time between CD4 counts being similar between DG recipients and non-recipients. The overall average rate of recovery per year among disability grant recipients and non-disability grant recipients was similar at an average of 62 cells/µL/year. The mean recovery rate for intervals which have a starting between 150 and 250 cells/µL is 145 cells/µL/year overall, and 126 cells/µL/year for intervals which are post-ART initiation.

## 1.2 Disability grants vs. CD4 recovery

Our empirical strategy has three components. The first component examines manipulation through CD4 count density difference across the threshold, noting that this threshold was also used for initiation of ART and as discussed below is inadequate evidence of manipulation. The second component is the main identification strategy. We utilize a difference-in-differences strategy on recovery rates, using the difference in the annualized CD4 rate of change between DG recipients and non-recipients when near and not near the qualification threshold. The third component tests the difference-in-difference model against false thresholds to ensure that the effect detected is specific to the theoretical qualification threshold of 200 cells/µL.

As a first step, we examine differences in the distribution of CD4 counts among disability grant recipients and non-recipients. If individuals were manipulating CD4 counts to qualify for disability grants, we would expect that there would be a discontinuous increase in the proportion of individuals with CD4 counts just under the qualification threshold of 200 cells/µL particularly for those who actually received disability grants. We formally test for discontinuity in the density of CD4 counts using density discontinuity tests by McCrary [29] and Cattaneo, Jansson [30]. However, given that ART guideline also recommend initiation at a CD4 threshold of 200 cells/µL [31, 32], it is possible that a larger number of individuals would appear in the system just below this level even without manipulation due to DG qualification rules.

Figure 1: Histogram of CD4 counts among disability grant recipients and non-recipients

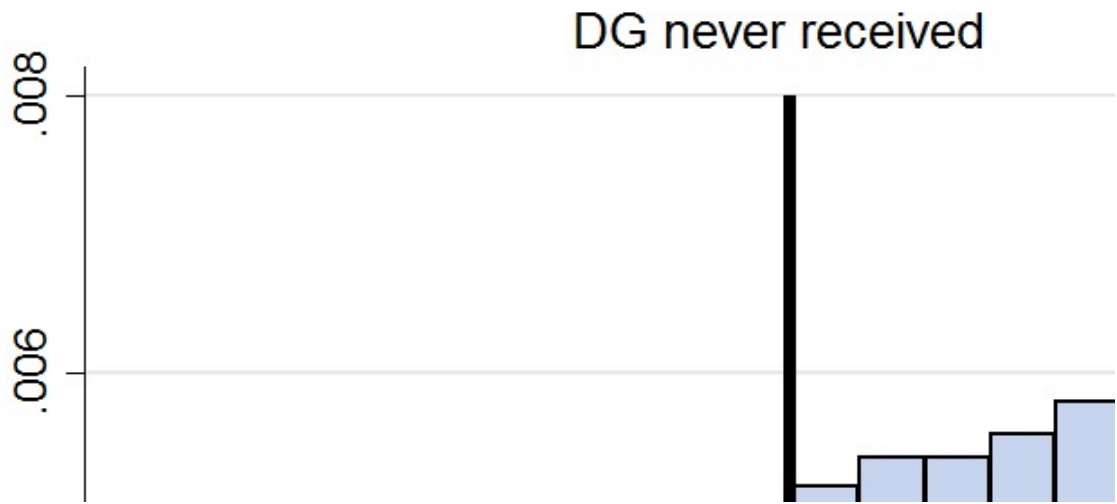

Disability grant recipients includes anyone in the dataset who had ever reported received a disability grant.

We note a small, though inconclusive excess mass in the distribution of CD4 counts under and just before the CD4 threshold of 200 cells/μL for those who have ever received disability grants in Figure 1, with no notable excess mass for those who have never received a DG. This is consistent with the possibility of CD4 count manipulation. However, neither the McCrary [29] nor Cattaneo, Jansson [30] excess mass tests are significant when testing formally.[2]

To address this potential selection concern, the main identification strategy used in his study utilizes a difference-in-differences (DID) strategy that focuses on recovery rates rather than levels of CD4 count. In the DID models estimated, the first compare general differences in CD4 recovery rates between DG recipients and DG non-recipients (first difference), and then estimate the relative difference in these two recovery rates around the CD4 cells/μL 200 cutoff (second difference). While we may expect that individuals who receive disability grants may differ in a variety of ways in terms of their general health trajectories from non-eligible individuals, we would not expect that these differences would be any different around the somewhat arbitrary CD4 cells/μL 200 cutoff. Any omitted factors which may threaten the causal validity of the difference-in-difference specification above are most likely to operate by changing the overall recovery rate over time, but not the differential recovery rate specifically at a CD4 threshold of 200 cells/μL.

---

[2] Both tests are performed using bandwidth bounds of 50 cells/μL around the 200 cells/μL threshold, otherwise using the default parameters and assumptions in the public Stata packages (http://eml.berkeley.edu/~jmccrary/DCdensity/ and the rddensity package, respectively). The population used was for disability grant recipients post initiation.

The main difference-in-difference model we estimate is given by:

Equation 1:
$$CD4recovery_{i,int} = \beta_0 + \beta_2 DGrecipient_i + \beta_3 threshold_{int} * DGrecipient_i + \beta_4 X_{i,int} + \sum \beta_{5j} initialCD4_{j,int} + \epsilon_{i,int}$$

where i is an individual, int is an interval between CD4 counts, CD4recovery is the annualized CD4 count change for the interval observation, threshold is a dummy variable for whether the starting CD4 count of the interval is close to the 200 cells/μL threshold using two main definitions (CD4 175-225 and 150-250 cells/μ), DGrecipient is whether the individual is recorded as ever receiving a disability grant, X is a vector of control variables (age, age$^2$, categorical dummies for years of education, sex, and categorical dummies for distance from nearest road), and initialCD4 is the CD4 count at the beginning of the recovery period expressed as a vector containing dummies for starting CD4 count values in bins of 25 cells/μL. A standalone variable for the threshold is unnecessary in this case, as all definitions of the threshold used are perfectly collinear with the initialCD4 dummies. In this case, $\beta_3$ is the variable of interest.

Figure 2: Annualized rate of CD4 bin

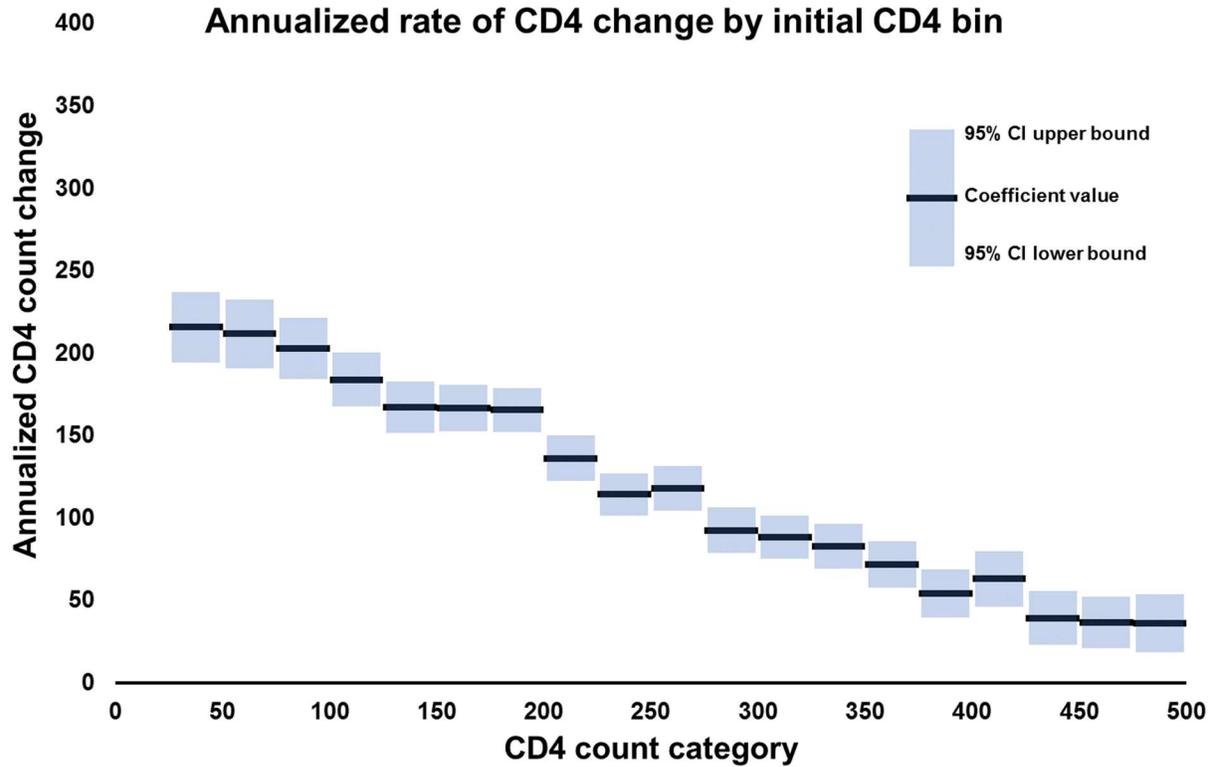

The coefficient values in this chart refer to the coefficients for a regression containing only dummies for each bin of width 25 cells/μL. This indicates the mean CD4 recovery rate for all observations with an initial CD4 in each of these bins

The fine dummy bins for initial CD4 count are necessary due to differences in CD4 recovery rate at difference starting points. There is an overall negative and relatively linear relationship between starting CD4 counts and CD4 recovery rate, as shown in Figure 3 (additional detail in Appendix 1). To control for the changes in the rate of recovery, we use the vector of initial CD4 count dummies in all regressions.

To visually inspect this difference-in-difference strategy, we run a regression model which interacts initial CD4 count bins of width 25 cells/μL with whether the individual had ever received a DG, as in below Equation 2. We then plot the value of the interaction coefficient and its 95% confidence interval for each of these bins.

Equation 2:

$$CD4recovery_{i,int} = \beta_0 + \sum_{b=1}^{21} \beta_{1b} startCD4bin_{b,int} + \beta_2 DGrecipient_i +$$
$$\sum_{b=1}^{21} \beta_{3b} startCD4bin_{b,int} * DGrecipient_i + \epsilon_{i,int}$$

Figure 3: Difference in annualized CD4 change by starting CD4 bins

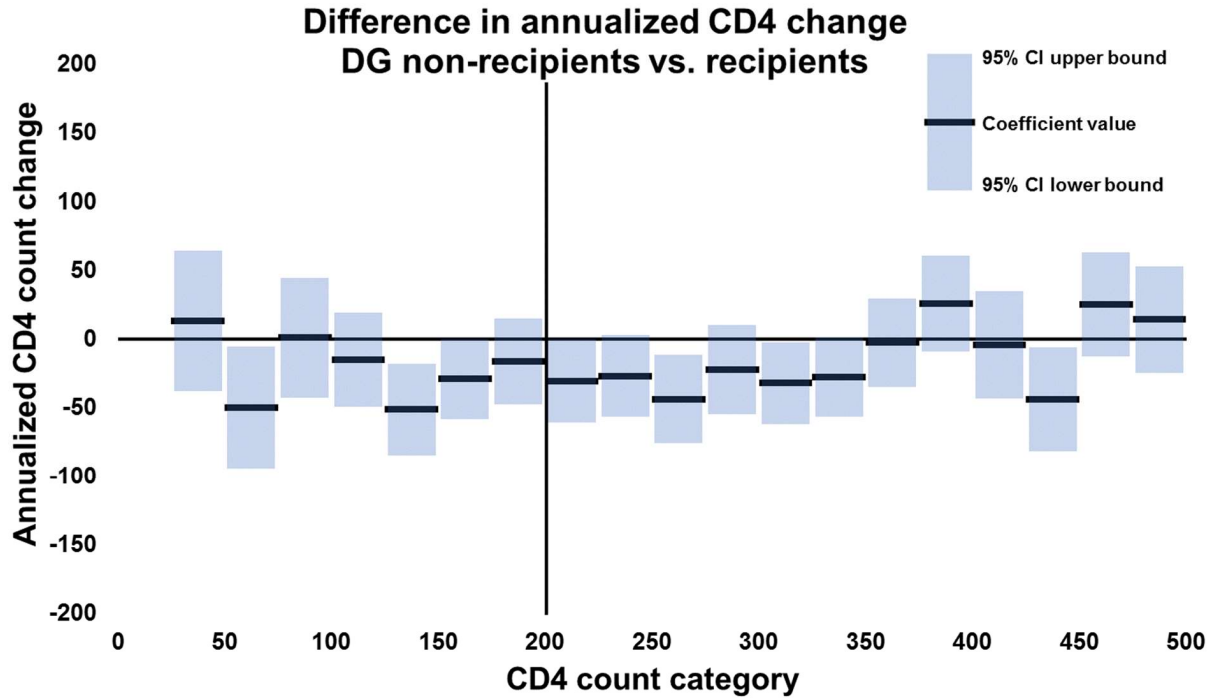

The coefficient values in this chart refer to the interaction terms in a regression based on Equation 2 between dummies for each bin of width 25 cells/μL and a time-invariant dummy of whether an individual was a disability grant recipient or not. Positive coefficients indicate that disability grant recipients with a starting CD4 in a given bin had higher recovery rates than non-recipients. Standard errors are clustered at the individual level.

Figure 3 shows estimated average difference in CD4 recovery rates between DG recipients and non-recipients. Disability grant recipients appear to have a slower overall rate of recovery than non-recipients at most CD4 levels near the threshold, although no notable difference in mean recovery rates is observed overall. This suggests that DG recipients and non-recipients are comparable populations with regard to CD4 recovery, and that differences between these populations are limited to the vicinity of the CD4 count threshold.

Table 2: Difference-in-Difference Estimates

| Dependent variable: Annualized CD4 change | | | | | | |
|---|---|---|---|---|---|---|
| Population: | All | All | All | Post-initiation | Post-initiation | Post-initiation |
| | (1) | (2) | (3) | (4) | (5) | (6) |
| Regression model: | OLS | OLS | OLS | OLS | OLS | OLS |
| **Disability grant received ever** | 12.1** | 14.6** | 18.2*** | 4.6 | 6.9 | 8.8 |
| | (5.8) | (6.0) | (6.2) | (6.4) | (6.5) | (6.8) |
| | | -31.0* | | | -33.2 | |
| Initial CD4 is between 175 and 225 * Disability grant received ever | | (17.4) | | | (21.9) | |
| Initial CD4 is between 150 and 250 * Disability grant received ever | | | -37.8*** (12.4) | | | -30.3** (15.1) |
| Initial CD4 count control vector | YES | YES | YES | YES | YES | YES |
| Year fixed effects | YES | YES | YES | YES | YES | YES |
| Demographic control vector | YES | YES | YES | YES | YES | YES |
| Constant | -89.5** (37.9) | -89.8** (37.9) | -90.2** (37.9) | -85.2** (43.4) | -85.5** (43.4) | -85.8** (43.4) |
| N (observations) | 24,509 | 24,509 | 24,509 | 14,825 | 14,825 | 14,825 |
| $r^2$ | 0.10 | 0.10 | 0.10 | 0.10 | 0.10 | 0.10 |

Notes: Individually-clustered standard error in parentheses. * p<0.10, ** p<0.05, *** p<0.01. Post-initiation refers to all observations which occur after an individual has initiated ART. Starting CD4 control vector includes dummies for starting CD4 bins of width 25 cells/µL. Demographic control vector includes age, age$^2$, categorical dummies for years of education, sex, and categorical dummies for distance from nearest road.

Table 2 shows a first set of DID results. We include a vector for starting CD4 counts, year fixed effects, and the vector of demographic and socioeconomic controls. We find that disability grant recipients have a lower rate of recovery for all starting CD4 band definitions, with coefficients generally negative and statistically significant for interactions with starting CD4 bands 175-225 cells/µL and 150-250 cells/µL, with a magnitude of -30.3 cells/µL/year (p = 0.035) when we consider only post-ART initiation in column (6).

Given that we observe most individuals multiple times, we also estimate a model with individual fixed effects. The within-person estimator is more robust against any inherent differences between disability grant recipients and non-recipients. Because all our measure of disability grant receipt (but not the interaction term) and all demographic covariates except age are time-invariant, these are dropped from the model.

Table 3: DID Results with Individual Fixed Effects

| Dependent variable: Dependent variable: Annualized CD4 change | | | | | | |
|---|---|---|---|---|---|---|
| Population: | All | All | All | Post-initiation | Post-initiation | Post-initiation |
| | (1) | (2) | (3) | (4) | (5) | (6) |
| Regression model: | FE | FE | FE | FE | FE | FE |
| Initial CD4 is between 175 and 225 * Disability grant received ever | 2.4 (12.4) | | | -9.7 (16.3) | | |
| Initial CD4 is between 150 and 250 * Disability grant received ever | | -4.4 (9.4) | | | -6.6 (12.6) | |
| First interval after having a CD4 > 200 | | | 24.1*** (5.9) -29.8*** | | | -26.5*** (6.8) -21.0* |
| First interval after having a CD4 > 200 * Disability grant received ever | | | (10.2) | | | (11.4) |
| Initial CD4 control vector | YES | YES | YES | YES | YES | YES |
| Year fixed effects | YES | YES | YES | YES | YES | YES |
| Demographic control vector | YES | YES | YES | YES | YES | YES |
| Constant | -1010.8*** | 44828 | 44828 | 26501 | 26501 | 26501 |
| | (132.0) | 8213 | 8213 | 6011 | 6011 | 6011 |
| N (observations) | 44,828 | 44,828 | 44,828 | 26,501 | 26,501 | 26,501 |
| Individuals | 8,213 | 8,213 | 8,213 | 6,011 | 6,011 | 6,011 |
| r² | 0.30 | 0.30 | 0.30 | 0.30 | 0.30 | 0.30 |

Individually-clustered standard error in parentheses. * p<0.10, ** p<0.05, *** p<0.01. Post-initiation refers to all observations which occur after an individual has initiated ART. Starting CD4 control vector includes dummies for starting CD4 bins of width 50 cells/µL and a polynomial term for starting CD4. Demographic control vector includes age and age$^2$.

The individual fixed effects models in Table 3 yield negative but largely insignificant results for the interaction terms of interest. As an alternative test, we use a dummy interaction for whether the observation starts with the first CD4 test after an individual goes over the qualification threshold in columns (3) and (6). This is the first time an individual will have been given a clear sign that he/she no longer qualifies by the CD4 threshold. These results show that disability grant recipients recover on average 21.0 cells/µL/year slower (p = .064) than non-recipients at this point (column 6).[3]

Our causal identification strategy relies on specificity to the CD4 count threshold of interest. Therefore, as both an identification strategy and as a robustness check, we systematically examine alternative false thresholds to observe whether there exists negative interaction terms between DG status and nearness to the threshold. We run the fixed effect regression in Table 3, column (5) for post-initiation data only, changing the treatment effect bin definitions to every tenth bin of width 100 cells/µL starting from 0 to 600 cells/µL cells/µL, matching the threshold width size from the main regression. These are the results of independently-run regressions, rather than multiple threshold coefficients in the same regression.

Figure 4: Treatment interaction coefficients with varying treatment bin definitions

---

[3] Results from the equivalent OLS specification (not shown) are similar, with a coefficient of -25.5 (p = .092)

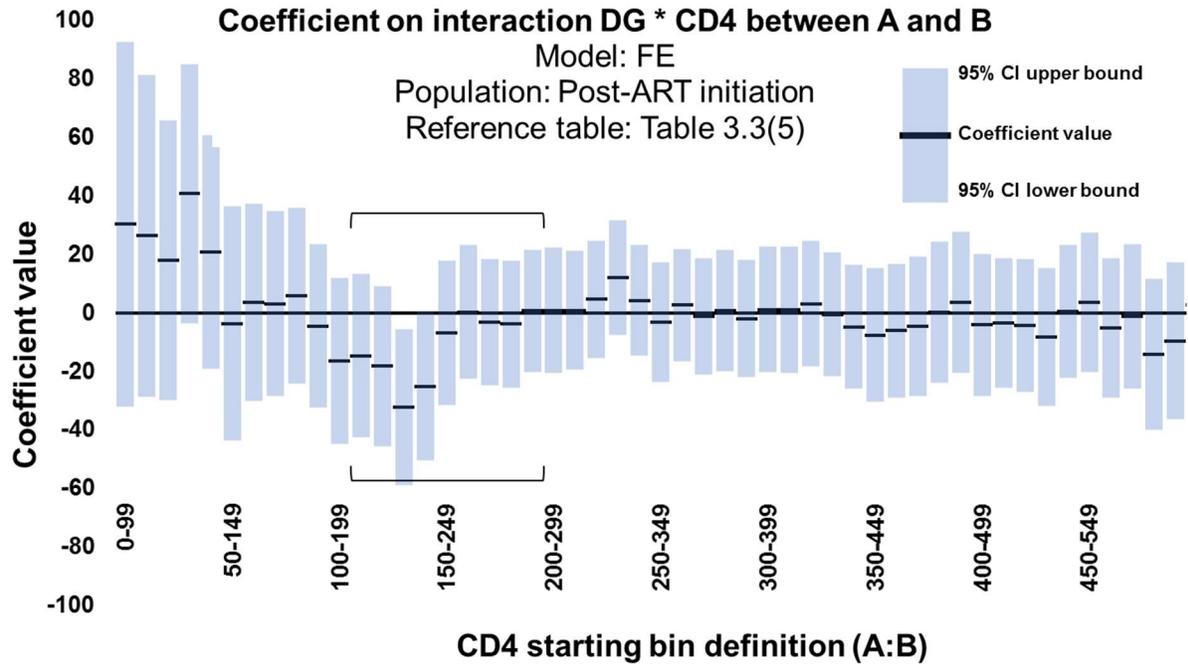

Each bar represents the value of the interaction term between the DG receipt dummy and the arbitrary threshold dummy, as defined in the x axis, for separate, independent regression runs. The bracketed area represents the regressions for which the threshold definition contains the actual DG qualification threshold. The regression model for each regression represented here is identical to that from Table 3, column 5, except for the threshold used.

The results of the post-initiation data, using our most conservative fixed effects model post-initiation, show precisely the pattern that would have been theoretically predicted in the presence of a threshold effect. Coefficients become negative when the treatment bin approaches and encompasses the treatment threshold, and are statistically significant at $\alpha = .05$ only when the treatment bin is starting CD4 levels of 130-229 cells/μL, with a difference-in-difference CD4 recovery estimate of -27 cells/μL/year (p-value = 0.035) slower around the threshold for disability grant recipients.

### 1.2.1 Robustness checks

As mentioned above, eligibility rules for the DG changes over time. Because the threshold effect should theoretically be stronger in years before the threshold was legally abolished, we take a sub-population of data in which both the starting and end dates of an interval are between and including 2003 to 2008, estimating disability grant ever status as 1 only if there is indication of receiving a disability grant between these dates. We note that population sizes are relatively small in this population, noting only 8,274 total intervals, of which 1,505 are from disability grant recipients, substantially limiting power. We do not find evidence that the law change had significant impact on behavior around the qualification threshold using only data from before the law change, nor do we find evidence that the threshold effect is stronger before the law change via the triple interaction, as shown in Appendix 2.

Figure 5: Treatment interaction coefficients with varying treatment bin definitions robustness checks

*Panel a:*

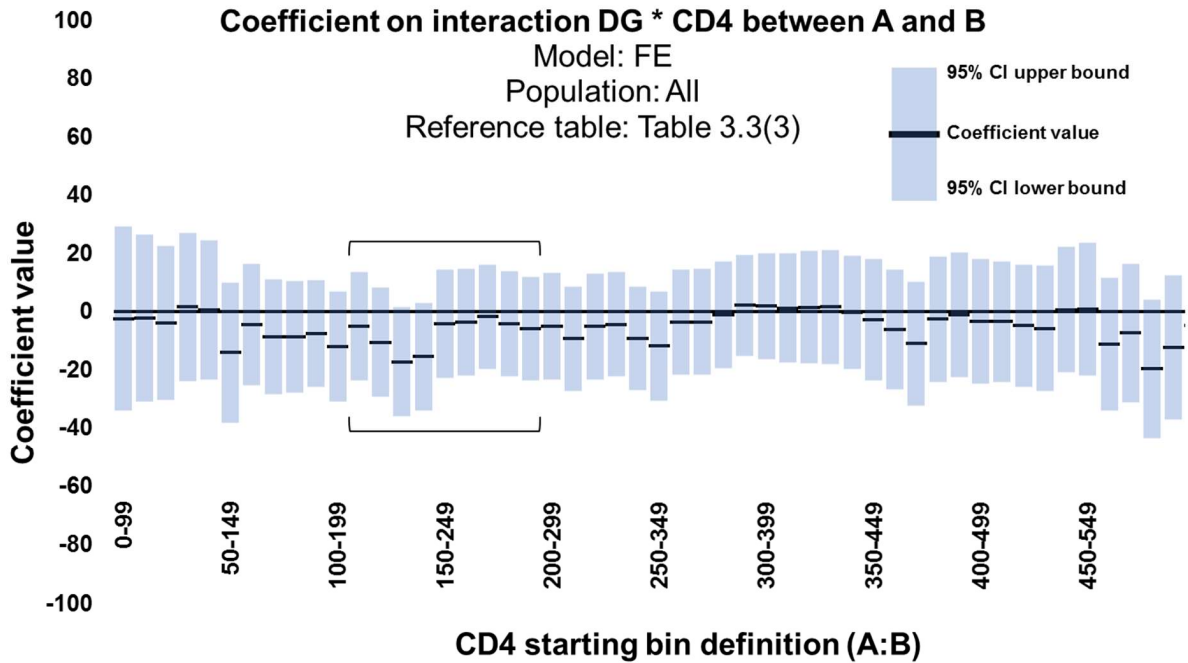

*Panel b:*

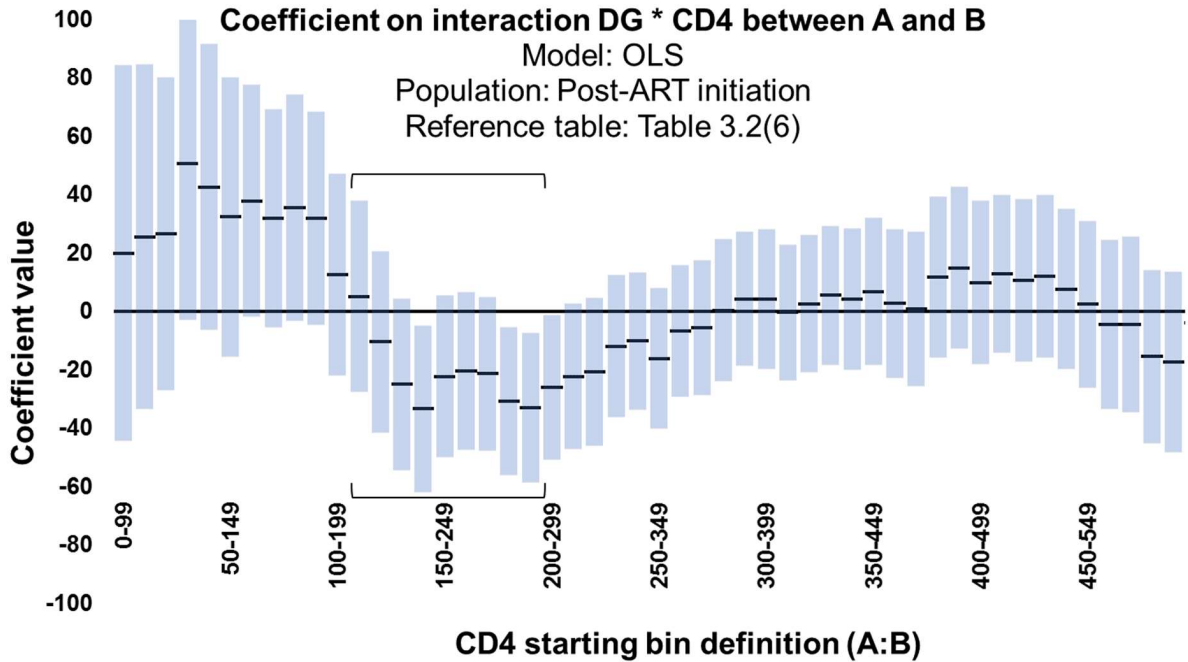

Each bar represents the value of the interaction term between the DG receipt dummy and the arbitrary threshold dummy, as defined in the x axis, for separate, independent regression runs. The bracketed area represents the regressions for which the threshold definition contains the actual DG qualification threshold. The regression model for each regression represented here is identical to that from Table 3, column 3 for *panel a* and Table 2, column 6 for *panel b*, except for the threshold used.

We include two additional falsification tests of the same type as Figure 4, changing the threshold of interest in separate regression models but using two alternative specifications of our regression results, as shown in Figure 5. The first uses the same statistical model, but for the entire population, while the second uses the OLS model from Table 2 for post-ART initiation. Both show a similar pattern of effect, where any differences between the recovery rates of disability grant recipients and non-recipients are negative and occur when the bandwidth definitions of interest approach the disability grant qualification threshold.

## 1.3 Disability grants vs. time to next CD4 count

Table 4: DID Estimates of DG Impact on Time Between Test

| Dependent variable: Time between CD4 count tests | | | | | | | |
|---|---|---|---|---|---|---|---|
| Population: | All | All | Pre-initiation | Post-initiation | Pre-initiation | Post-initiation |
| | (1) | (2) | (3) | (4) | (5) | (6) |
| Regression model: | OLS | FE | FE | FE | FE | FE |

| | (1) | (2) | (3) | (4) | (5) | (6) |
|---|---|---|---|---|---|---|
| Disability grant received ever | -3.7 | | | | | |
| | (3.7) | | | | | |
| Initial CD4 is under 200 cells/µL * Disability grant received | | 17.0*** | | | | |
| | | (5.6) | | | | |
| Initial CD4 is 200-250 cells/µL * Disability grant received | | | -26.9 | 7.1 | | |
| | | | (20.7) | (8.8) | | |
| First interval after having a CD4 > 200 * Disability grant received ever | | | | | -31.6*** | -19.6*** |
| | | | | | (3.8) | (3.9) |
| First interval after having a CD4 > 200 * Disability grant received ever | | | | | -11.1 | -2.2 |
| | | | | | (7.4) | (7.1) |
| First interval after having a CD4 > 200 * Starting CD4 between 200 and 250 * Disability grant received ever | | | | | 39.7*** | 25.6** |
| | | | | | (12.4) | (12.0) |
| Initial CD4 count control vector | YES | YES | YES | YES | YES | YES |
| Year fixed effects | YES | YES | YES | YES | YES | YES |
| Demographic control vector | YES | YES | YES | YES | YES | YES |
| Constant | 318.6*** | 10012.1*** | 10940.7*** | 6637.5*** | 9995.0*** | 6657.3*** |
| | (20.5) | (131.1) | (379.9) | (151.2) | (136.0) | (151.2) |
| N (observations) | 24,997 | 45,729 | 6,470 | 26,905 | 45,729 | 26,905 |
| Individuals | - | 8,221 | 2,976 | 6,021 | 8,221 | 6,021 |
| r² | 0.05 | 0.4 | 0.5 | 0.3 | 0.4 | 0.3 |

Individually-clustered standard error in parentheses. * $p<0.10$, ** $p<0.05$, *** $p<0.01$, - variable omitted for due to perfect collinearity. Pre-/Post-initiation refers to all observations which occur before/after an individual has initiated ART. Starting CD4 control vector includes dummies for starting CD4 bins of width 25 cells/µL and a polynomial term for starting CD4. Demographic control vector includes age, age$^2$, categorical dummies for years of education, sex, and categorical dummies for distance from nearest road for OLS models, and age and age$^2$ for FE models.

We consider two main hypotheses in this section. First, we consider whether being currently qualified for a disability grant causes recipients to more regularly receive laboratory tests. Because re-application occurs every 6 months, this gives disability grant recipients incentive to have CD4 counts more frequently. If this were the case, we would expect that disability grant recipients would have shorter times between clinic visits while they are still qualified. To test this, we utilize difference-in-difference approach as above, but using being currently qualified (i.e. having a starting CD4 count of <200 cells/µL) as the second difference instead of proximity to the qualification threshold.

While the DG recipients overall have an average of 24.3 ($p<0.001$) fewer days between visits, this difference disappears when controlling for demographic differences as shown in column (1) of Table 4. DG recipients have on average 17 more days ($p<0.001$) between CD4 counts, suggesting that the DGs do not increase the frequency of tests for qualifying participants in column (2).

The second hypothesis tested is manipulation by "probabilistic shopping." Individuals who have a CD4 count that is immediately above the qualification threshold are not qualified for the grant, but are close enough that another test may put them under the threshold due either to declining health or random variation in the test itself. In that case, those individuals just over the threshold would want to go back to the clinic and have another CD4 count sooner than otherwise. In this case, the differenced region of interest is having a CD4 count of 201-250 cells/µL.

We find no evidence of "probabilistic shopping" for either pre- or post-initiation in columns (3-4) of Table 4. However, the triple interaction between DG receipt, first over the threshold, and if the observation was relatively close to the threshold are positive and highly statistically significant in columns (5-6), suggesting a possible disappointment effect, where individuals are less likely to return to the clinic after learning of loss of grant status.

# Discussion

The analysis here presents evidence that the structure of the qualification rules of South Africa's disability grant program unintentionally caused a slowing of CD4 recovery for patients close to the qualification threshold in rural KwaZulu-Natal, South Africa. This threshold effect appears relatively robust against model specification and falsification tests, including falsification by alternative threshold window definitions. The reduction in recovery found was 20-30 CD4 cells/µL/year, which translates to an approximate 20% reduction in the average recovery rate over the impacted population. We hypothesize that the threshold effect on CD4 recovery observed here is most likely driven by changing ART adherence. Alternative causes of this slowdown, such as record manipulation, are unlikely, as the data here are directly from clinical laboratory records that are hard to access for patients. The identification strategy used in this analysis detected manipulation in a scenario where more conventional tests, such as excess mass tests, were not appropriate, and is potentially applicable to other thresholds in which longitudinal data are available.

The identification strategy in this paper is unusual, but potentially powerful. We identify the existence of manipulation of CD4 counts with relatively minimal data – our main models requiring only an individual identifier, CD4 test results, and dates – but a very limited sense of what the size of the effect might be in meaningful terms. In the specific circumstances of the outcome in question, there are potentially outsized and wide-reaching consequences of even a small effect may be enough for specific policy implications to address them, as discussed below. However, in other circumstances, merely demonstrating that a manipulation effect exists without comparison to the overall impact of the program may not be appropriate.

Two key limitations in this study are the lack of direct adherence data and specific dates of disability grant application. We detect the downstream consequences of patient behaviors through CD4 counts, but do not observe the behaviors themselves. Adherence data, whether through refill dates or other means, would allow more direct testing of the proposed causal mechanism. However, we observe the most relevant downstream consequence of lack of adherence. Furthermore, alternative explanations for the decline in CD4 counts specific to DG recipients at the threshold are either also likely due to declines in health due to the threshold or not plausible with our data. If individuals were able to reduce their CD4 counts by means other than reduced adherence, that is still a substantial health issue due to the threshold effect. Secondly, manipulation through falsification of paperwork is unlikely to appear in our tests, as our data are from the laboratories rather than the social security offices or application forms.

Similarly, because grant reapplication is due every six months, we would expect threshold effects to be strongest shortly before reapplication. Unfortunately, neither of these types of data are available at the time of this writing. However, the net effect of the lack of disability grant application dates and imprecision of grant receipt status is most likely bias towards the null, resulting in our standard errors being overly conservative and unlikely to detect effects. The lack of this data could be thought of either as a spillover or cross-contamination effect, where there may be individuals marked as being a DG recipient who have it for reasons not related to HIV and/or individuals who are marked as being non-recipients who did not report their DG status, but did receive support. Alterively, we could consider it measurement error in our independent variable, which also generally yields bias towards the null.

While we find that the program causes negative health behaviors specific to the CD4 qualification threshold, the effect of this incentive must be weighed against the net benefits to recipients. The disability grant allows poor individuals some minimal level of income and helps ensure food availability, which is additionally important for tolerating ART, and as such is likely to be a substantial net benefit to recipients, even with this negative disincentive. Recent modelling work by Low and Pistaferri [33] suggests that even with many of the known disincentive effects, increasing disability benefits may be a net welfare increase.

Several policy considerations may be able to eliminate, or even reverse, the incentives provided by the DG, without reducing the other welfare benefits associated with the program. Firstly, clearer guidelines for grant qualification should be provided to clinical workers in the absence of the CD4 threshold rule. Secondly, South Africa should lengthen the amount of time which the grant covers from six months to two years, allowing time for individuals to recover and return to the workplace as in Bor, Tanser [34], potentially reducing reliance on public grants. Thirdly, disability benefits for treatable conditions, such as HIV/AIDS could be single grants with no reapplication allowed for the same illness, especially if the grant period is lengthened to two years. There is little incentive to manipulate CD4 recovery if recipients do not requalify for grants. Finally, alternative benefit structures which reverse the direction of the incentive at the qualification threshold could be considered. This could include tying grant status to adherence through refill monitoring, where successful reapplication requires a good adherence record. Alternatively, the grant could be redesigned such that grant amounts increase for achieving higher CD4 counts for some fixed period of time and possibly subsequently taper as a person becomes health enough to enter the labor force.

Though the circumstances that resulted in negative health manipulation are, in this case, specific to the South African disability grants program, we can hypothesize that this type of manipulation is broadly more likely to occur if: 1) qualification rules rely on a known quantitative indicator of disability, 2) the potential recipient is able to change their behaviors to move the value of the indicator, 3) the monetary benefits of disability pay are high compared to outside options, 4) the perceived utility from those monetary benefits are high compared to the perceived utility of the relevant positive behaviors, and 5) levels of future discounting against the negative consequences of manipulation are high. While these conditions do not differ substantially from general principles for hardship relief policies, our paper adds evidence that these incentives may have consequences for health. We could further theorize that these health consequences could, in turn, further reduce labor supply in the long run. As low- and middle-income countries, particularly those in sub-Saharan Africa, continue to experience rapid economic growth and consider implementing additional hardship-related benefits policies, care will need to be taken to ensure that the incentives they create do not undermine the policies' intended results.

Appendix 1: Annualized CD4 count change over starting CD4 levels

*The figures below were generated using coefficients on the starting CD4 bin dummies, using the following regression:*

$$CD4recovery_{i,int} = \beta_1 startCD4_{0:24,int} + \beta_2 startCD4_{25:49,int} + \beta_3 startCD4_{50:74,int} + \ldots + \beta_{21} startCD4_{\geq 500,int} + \epsilon_{i,int} = \sum_{b=1}^{21} \beta_b startCD4bin_{b,int} + \epsilon_{i,int}$$

where CD4recovery is the annualized rate of CD4 count change between two CD4 count intervals, $startCD4bin_{1,int}$ refers to the dummy that corresponds to having a CD4 count between 0 and 24 cells/µL, $startCD4bin_{2,in}$ corresponds to the starting CD4 between 25 and 49 cells/µL, and so on.

Annualized rate of CD4 change by starting CD4 category

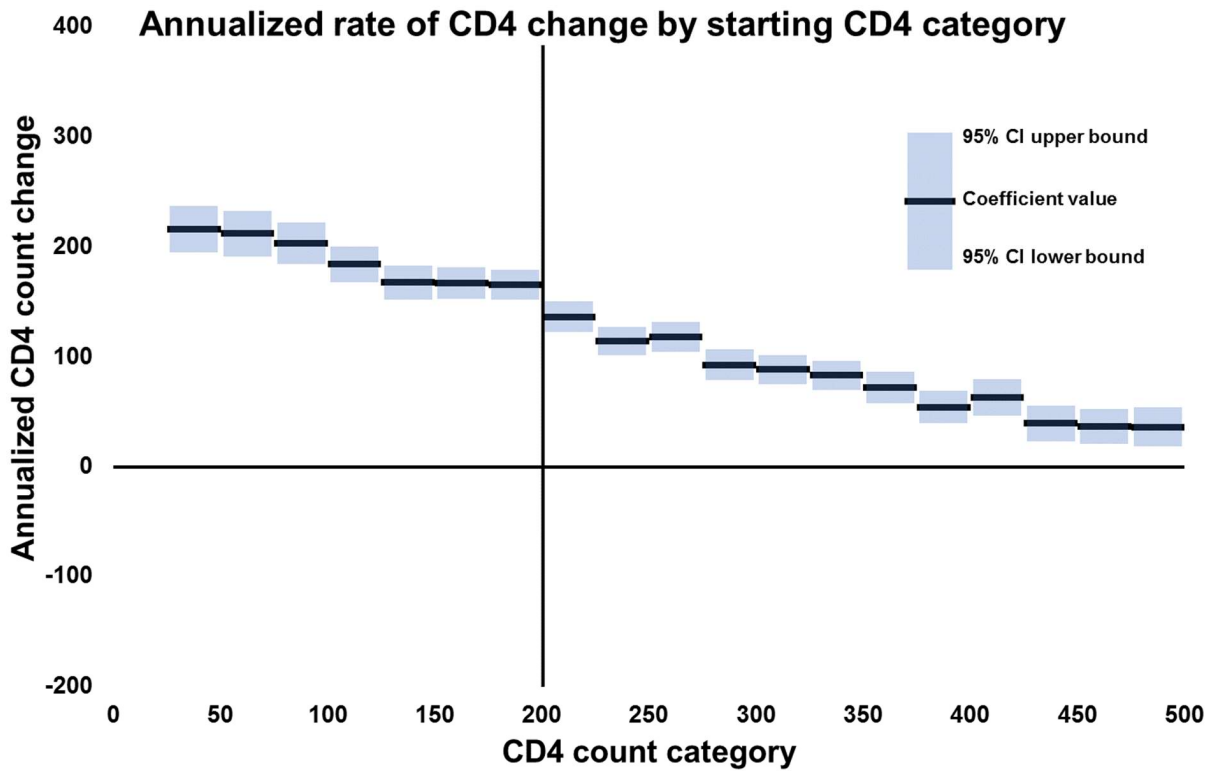

As shown above, the rate of annual CD4 count change is highest among those who start each interval with the lowest CD4 counts, averaging a rate of 215 cells/µL per year (95% CI: 195-237) among those whose CD4 to at the start of the interval was between 25 and 49 cells/µL. This rate of improvement gradually tapered toward higher CD4 counts, appearing to converge to 0 well after reaching a CD4 count level of 500 cells/μL, reaching "normal" levels of CD4 counts.

Annualized rate of CD4 change by starting CD4 category, pre- vs. post-ART initiation

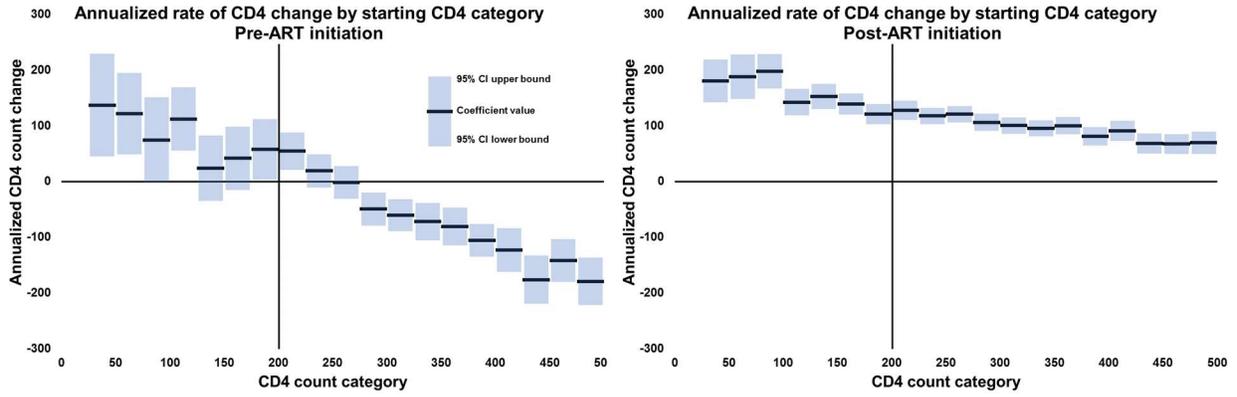

As expected, the average rate of annualized CD4 change from a given CD4 count level is generally negative before ART initiation and positive post-initiation as shown above.

Appendix 2: Robustness check: Pre- vs. post- Law change in 2008

Dependent variable: Annualized CD4 change

| | Population: | All | All | All | Post-initiation | All | All |
|---|---|---|---|---|---|---|---|
| | | (1) | (2) | (3) | (4) | (5) | (6) |
| | Model: | OLS | OLS | FE | FE | OLS | FE |
| **Disability grant received ever** | | -4.8 | | | | 12.2 | |
| | | (3.7) | | | | (7.9) | |
| **Disability grant received pre-law change (2003-2008)** | | 4.4 | -5.4 | | | 11.4 | |
| | | (4.5) | (8.1) | | | (10.1) | |
| **Starting CD4 between 150 and 250 * DG received pre-law change** | | | | 14.7 | 56.8 | | |
| | | | | (27.6) | (47.8) | | |
| **Starting CD4 is between 150 and 250 * Disability grant received ever** | | | | | | -38.2** | -12.0 |
| | | | | | | (15.5) | (11.8) |
| **Starting CD4 between 150 and 250 * DG received pre-law change * Disability grant received** | | | | | | 1.4 | 19.1 |
| | | | | | | (21.1) | (16.1) |
| **Starting CD4 control vector** | | | | YES | YES | YES | YES |
| **Year fixed effects** | | | | YES | YES | YES | YES |
| **Demographic control vector** | | | | YES | YES | YES | YES |
| **Constant** | | 62.4*** | 87.6*** | 2422.2*** | 2636.9*** | 164.8*** | 907.7*** |
| | | (1.4) | (4.4) | (566.1) | (803.5) | (45.2) | (138.3) |
| **N (observations)** | | 45,921 | 6,397 | 6,197 | 2,686 | 22,584 | 40,854 |
| **Individuals** | | - | - | 2,746 | 1,178 | - | 7,202 |
| **r²** | | 0.00 | 0.00 | 0.00 | 0.00 | 0.00 | 0.00 |

Individually-clustered standard error in parentheses. * p<0.10, ** p<0.05, *** p<0.01. Post-initiation refers to all observations which occur after an individual has initiated ART. Starting CD4 control vector includes dummies for starting CD4 bins of width 50 cells/μL and a polynomial term for starting CD4. Demographic control vector includes age and age$^2$.

Appendix 3: Distribution of CD4 recovery rates by starting CD4 category

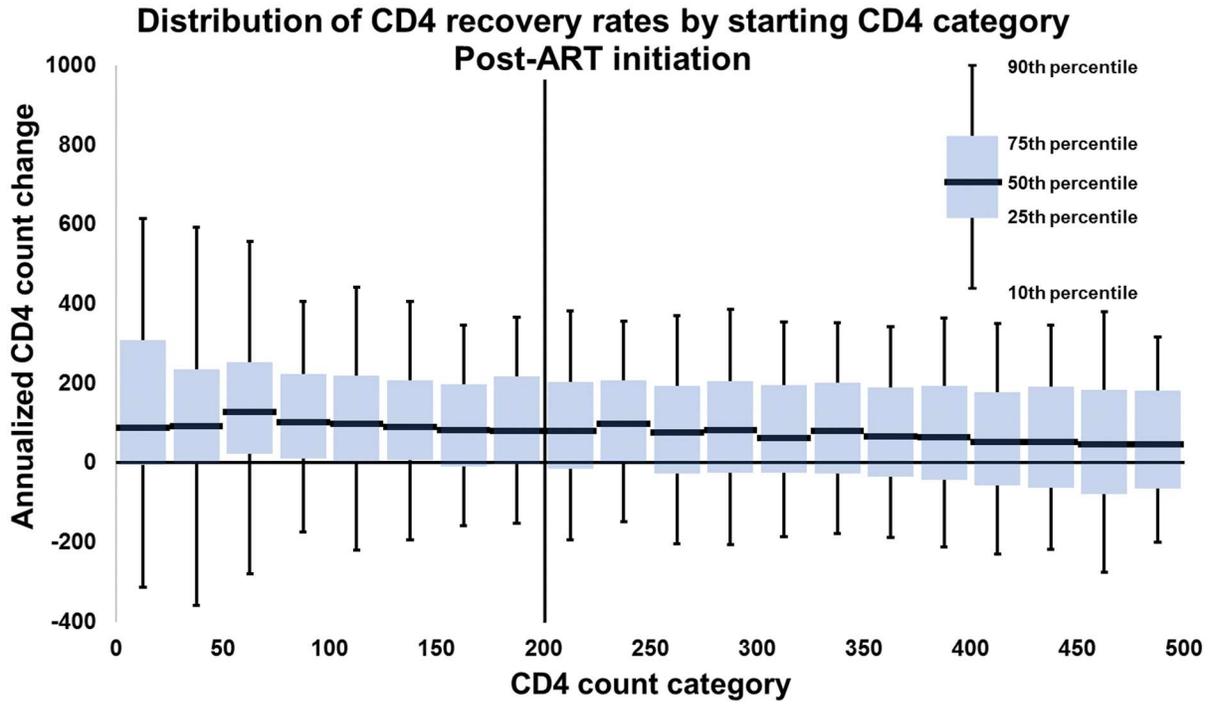

The bars above represent the distribution of CD4 recovery rates among those with a CD4 count at the start of the interval in each of the bins of width 25 cells/μL. The black bar represents the median recovery rate, the extent of the bars is the interquartile range, and the extent of the brackets is the 90$^{th}$/10$^{th}$ percentile limits.

Appendix 4: Difference time between CD4 counts by starting CD4 bins, pre- and post-ART initiation

*Panel a:*

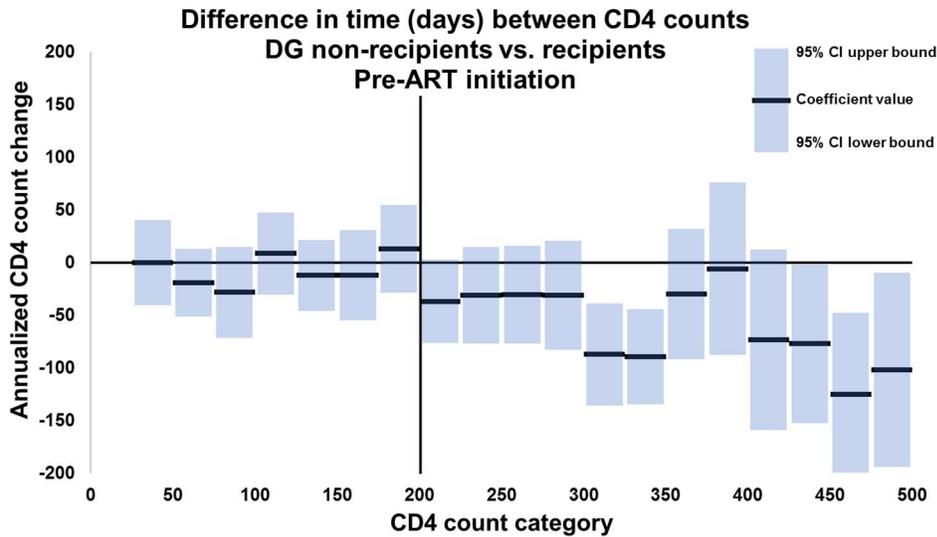

*Panel b:*

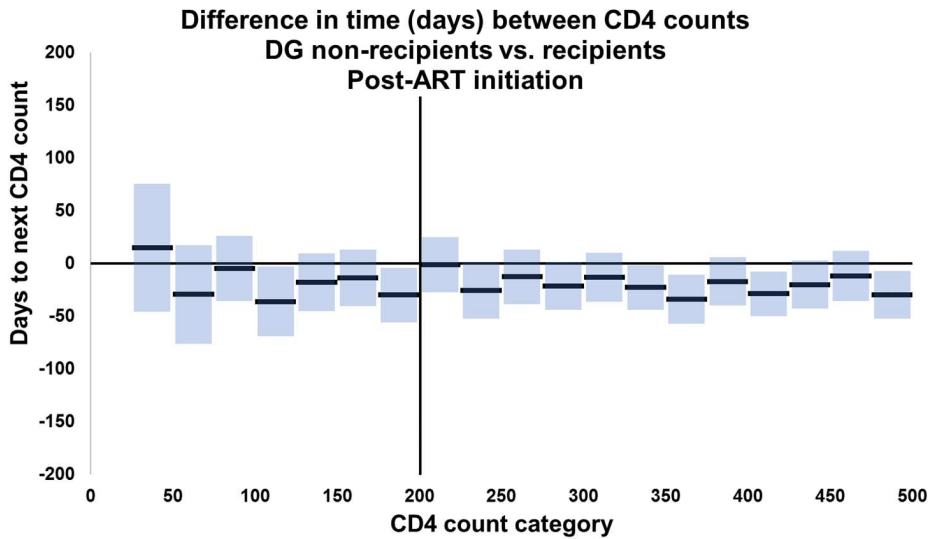

The coefficient values in this chart refer to the interaction terms in a regression based on Equation 2 between dummies for each bin of width 25 cells/µL and a time-invariant dummy of whether an individual was a disability grant recipient or not. Positive coefficients indicate that disability grant recipients with a starting CD4 in a given bin had more time between CD4 counts than non-recipients. Standard errors are clustered at the individual level.